\documentclass[preprint,12pt]{elsarticle}

\usepackage{amssymb}
\usepackage[tight, scriptsize]{subfigure}
\usepackage[]{captcont}
\usepackage{tikz}
\usepackage{tipa}
\usepackage{tabularx}

\journal{Biomedical Signal Processing \& Control}

\begin{document}

\begin{frontmatter}



\title{A Transversal Study of Fundamental Frequency Contours in Parkinsonian Voices}

\author[CITSEM]{Pablo Rodr\'{\i}guez P\'erez}
\author[CITSEM]{Rub\'en Fraile}
\author[HS,CEU]{Miguel Garc\'{\i}a-Escrig}
\author[CITSEM]{Nicol\'as S\'aenz-Lech\'on}
\author[CITSEM]{Juana M. Guti\'errez-Arriola}
\author[CITSEM]{V\'{\i}ctor Osma-Ruiz}
\address[CITSEM]{Centro de Investigaci\'on en Tecnolog\'{\i}as Software y Sistemas Multimedia para la Sostenibilidad\\ 
	Universidad Polit\'ecnica de Madrid - Campus Sur\\
	C/ Alan Turing 3. 28031 Madrid (Spain)}
\address[HS]{Hospital de Sagunto\\
	Av. Ram\'on y Cajal, s/n. 46520 Sagunto, Valencia (Spain)}
\address[CEU]{Medicine Department\\
	Universidad CEU Cardenal Herrera\\
	C/ Grecia 31. 12006 Castell\'on (Spain)}

\begin{abstract}
A transversal study of the pitch variability of parkinsonian voices in read speech is presented. 30 patients suffering from Parkinson’s disease (PD) and 32 healthy speakers were recorded while reading a text without voiceless phonemes. The fundamental frequency contours were calculated from the recordings, and the following measures were used for describing them: mean, minimum, maximum, and standard deviation of the estimated fundamental frequencies. Results based on these measures indicate that the influence of PD on some aspects of intonation can be masked by the effects of aging, especially for male voices. However, some parameters such as the relative fundamental frequency range exhibit lower correlations with age than with PD stage, as evaluated using the Hoehn and Yahr scale. These correlations between relative fundamental frequency range and PD stage reach moderate-to-high values in the case of women. 

Additionally, three parameters describing the form of the fundamental frequency modulation spectrum were investigated for correlation with age and PD stage. The study of this modulation spectrum provides some insight into the ability of the speakers to plan the intonation of full phrases. For both male and female populations, significant correlations were found between parameters obtained from the modulation spectrum of fundamental frequency and the PD stage. Nevertheless, the quantitative assessment of the performance of regression models built from these modulation parameters and fundamental frequency range suggests that such measures are likely to be of limited value in the early diagnosis of PD due to inter-speaker variability. 
\end{abstract}

\begin{keyword}
Parkinson's disease \sep Voice analysis \sep Fundamental frequency \sep Correlation analysis \sep Computer aided diagnosis 
\end{keyword}

\end{frontmatter}


\section{Introduction}
\label{}

Parkison's disease (PD) is one of the most frequent movement disorders affecting the eldest population. Its prevalence has been estimated to be between 1\% and 2\% for the population over 65, while this figure grows to between 3\% and 5 \% for people over 85 \cite{AFPG08}. In absolute numbers, its prevalence in the world was between 4.1 and 4.6 million in 2005 \cite{WACT11}. Due to differences in life expectancy, PD is more common in developed countries, though it is found in all ethnic groups \cite{AFPG08}.

Between 70\% and 89\% of PD patients report vocal difficulties \cite{MHAA05}. These difficulties have been classified into four different classes, according to their nature \cite{MHAA05,Crit81}: respiratory, phonatory, articulatory and prosodic. Hypophonia probably is the most widely recognized vocal impairment associated to PD \cite{Jank08}; it is mainly related to respiratory and phonatory functions. Disordered prosody is likely to be the second most relevant vocal impairment of parkinsonian speakers \cite{RCRR11}. It includes symptoms such as monotony \cite{Jank08}, speech rate abnormalities \cite{BGOB15}, difficulties in initiating speech and finding words \cite{Jank08}, and abnormal speech rate \cite{BGOB15,SGMS13,Skod15}.

Regarding intonation, the key difference between PD patients and age-matched healthy speakers seems to be the narrowing of the pitch range in PD patients, as detected in several studies \cite{RCRR11,SGMS13,INSH15}. Cnockaert \emph{et al} also found relevant differences in pitch modulation rates \cite{CSAO08}. The effect of PD on pitch range and intonation was found to be sex-specific by MacPherson \emph{et al} \cite{MaHS11}. Consistently, Ikui \emph{et al} also detected differences between sexes in the way pitch range was reduced \cite{INSH15}: lowering of the highest fundamental frequency ($f_\mathrm{o}$) happened for both sexes, while elevation of the lowest $f_\mathrm{o}$ was significant only for males. Furthermore, Skodda \emph{et al} found pitch range reduction to be more significant in females than in males \cite{SGMS13}.

The relation between PD and speech impairment has led to conjecturing about the potential use of speech (or voice) analysis as a cue for the early diagnosis of PD. Nevertheless, Becker \emph{et al} pointed out that tests measuring alterations in speech yielded results exhibiting only a limited sensitivity to PD, and also suffered from low specificity \cite{BMBB02}. Midi \emph{et al} \cite{MDKC08} did not find strong relations between voice impairment and the motor component of UPDRS (Unified Parkinson's Disease Rating Scale) either \cite{MDS03}. However, speaking involves both motor and cognitive activities \cite{Crit81} and, while motor disorders may not appear at the onset of PD, cognitive function assessment via speech analysis may help to identify subjects suffering from the illness at early stages \cite{Adle11}. The fact that intonation disorders in parkinsonian speakers cannot be fully explained by motor impairments \cite{MaHS11} implies the influence of cognitive decline in such disorders, hence suggesting the prospective usefulness of intonation analysis in the early diagnosis of PD \cite{HaCS04}. The relation between speech prosody (including intonation) and cognitive decline in PD patients has been investigated recently with positive results \cite{RMJK16}. In this context, the interest of speech analysis as a potential tool for the early diagnosis of PD has been renewed \cite{HCTS17}.

In this paper, we present a transversal study in which the intonation corresponding to a text read by 30 PD patients and 32 healthy speakers was analysed by estimating the corresponding fundamental frequency contours. Specifically, the following measurements of the fundamental frequency range were obtained: mean, minimum, maximum, and standard deviation. In addition, the spectra of the fundamental frequency contours were described by analysing the distribution of its energy among several intervals of modulation frequencies within the 0-20 Hz range. We have aimed at contributing to assess the value of intonation analysis as a diagnostic tool for PD. This requires transversal studies which incorporate the inter-speaker variability that actually happens in the clinic. The stage of PD corresponding to each patient was evaluated by means of the Hoehn and Yahr (H\&Y) scale \cite{HoYa67}. We have carried out an analysis of correlations between intonation descriptors and H\&Y stage previously to studying the discriminative capacities of these vocal features in classifying voices as belonging to either PD patients or healthy individuals. 

\section{Materials}
\label{sec:Materials}
30 outpatients of the Neurology Service at the Hospital de Sagunto were recorded between April and November 2015. The recording protocol was approved by the ethics committee of the Hospital and all participating patients signed informed consent before being recorded. They were recorded in a quiet room within the hospital, after seeing the neurologist. Background noises coming from contiguous rooms could not be avoided, but it was checked that they did not affected the results of the voice processing algorithms applied to the recorded signals. Before recording, the neurologist collected a data sheet for each patient, including information on sex, age, years since PD was diagnosed for the first time, and illness evolution stage according to the H\&Y scale. 

The group of patients included 11 women and 19 men, with assigned H\&Y labels ranging from 1 to 4. The average age of men was 74.1 years (standard deviation, $\sigma=10.2$), while for women the average was 72.2 years ( $\sigma=7.9$). More detailed information on the distribution of patients by sex, age and H\&Y stage is summarised in Table \ref{tab:Speakers}. Among the 30 patients, 4 suffered from tremoric PD (i.e. tremor was the dominant clinical symptom) and 1 suffered from akinetic PD. As for the time since PD was firstly diagnosed, at the time of recording it was 8.6 years on average for men ( $\sigma=5.1$), and 9.6 years on average for women ( $\sigma=5.1$).

A control group formed by 32 volunteers (12 women, 20 men) with no previously diagnosed neurological disorders were later recorded in similar conditions as the patients, and using the same equipment. Information about non-neurological disorders (e.g. laryngeal) potentially affecting prosody was not collected. Volunteers were given the same information as the patients, and they all signed informed consents too.  The average age of men was 70.0 years ( $\sigma=11.0$), while for women the average was 71.5 years ( $\sigma=9.0$). Table \ref{tab:Speakers} also includes more details on the distribution of the control group by sex and age.

\begin{table}
	\centering
	\begin{tabular}{|c|c|c|c|c|c|c|c|c|c|c|}
		\hline
		& \multicolumn{10}{|c|}{\textbf{Hoehn \& Yahr stage}} \\
		\cline{2-11}
		\textbf{Age} & \multicolumn{2}{|c|}{\textbf{0 (Control)}} & \multicolumn{2}{|c|}{\textbf{1}} & \multicolumn{2}{|c|}{\textbf{2}} & \multicolumn{2}{|c|}{\textbf{2.5 - 3}} & \multicolumn{2}{|c|}{\textbf{4}} \\
		\cline{2-11}
		& \textbf{M} & \textbf{F} & \textbf{M} & \textbf{F} & \textbf{M} & \textbf{F} & \textbf{M} & \textbf{F} & \textbf{M} & \textbf{F} \\
		\hline
		\textbf{50-59} & 4 & & & & 2 & & & & & \\
		\hline
		\textbf{60-69} & 6 & 5 & & & 3 & 3 & & 1 & & \\
		\hline
		\textbf{70-79} & 6 & 5 & 2 & 2 & 3 & 2 & & 1 & 1 & \\
		\hline
		\textbf{80-89} & 3 & 2 & 2 &  & 3 &  & 3 & 1 & & 1\\
		\hline
		\textbf{$\ge$90} &1 & & & & & & & & & \\
		\hline
	\end{tabular}
	\caption{Distribution of speakers by sex (F stands for females; M, for males), age and H\&Y label. The H\&Y scale was originally designed with only integer values, while a later modification introduced intermediate stages 1.5 and 2.5 \cite{GPRS04}. Although the neurologist used the original scale, one female patient was labelled as ``between grades 2 and 3''. This patient was assigned label 2.5 for the analysis, and grouped with grade 3 patients in this table for compactness. Control speakers have been assigned the H\&Y label 0.}
	\label{tab:Speakers}
\end{table}

The equipment used for recording consisted of a microphone, a mixer, and a personal computer (PC). A lavalier microphone was chosen in order to maintain the distance between mouth and microphone as fixed as possible, while avoiding the additional stress that head-mounted microphones might have caused to some patients. Namely, a \emph{Fonestar FCM-410} was selected due to its bandwidth: 30 Hz to 18000 Hz. A \emph{Fonestar SM-303SC} mixer was used for amplifying the microphone signal and directing it to the USB port of the PC. The open-source software \emph{Audacity} was run in the PC to manage analogue-to-digital conversion. This was performed at 44100 samples per second and 16 bits per sample.

All participants were requested to read a sentence in Spanish containing only voiced phonemes, and including the five Spanish vowels: /\emph{\textipa{'a \textgamma wa \dh e '\textlambda u \textbeta ja '\textbeta e \textbeta o i 'la \textbeta o en el 'la \textgamma o}}/. They were asked to perform reading at comfortable pace and intensity. The time taken for the reading task ranged from 2.35 s to 7.08 s, except for one patient (male, aged 84) that had some reading difficulties and needed 25.8 seconds, during which he performed several word repetitions. Table \ref{tab:Times} provides more details on the distribution of reading task durations and phonation times.

\begin{table}
	\centering
	{\small 
	\begin{tabular}[b]{|c|*6{p{1.3cm}|}}
		\hline
		 & \multicolumn{2}{p{2.6cm}|}{\centering \textbf{Reading task duration (s)}} 	& \multicolumn{2}{p{2cm}|}{\centering \textbf{Phonation time (s)}} 	& \multicolumn{2}{p{2cm}|}{\centering \textbf{Phonation time (\%)}} \\
		\cline{2-7}
		\textbf{Group} & $\mu$ &  $\sigma$ & $\mu$ &  $\sigma$ & $\mu$ &  $\sigma$ \\
		\hline
		\hline
		\textbf{Patients} & $\mathbf{4.65}$ &  $\mathbf{4.20}$ & $\mathbf{2.83}$ &  $\mathbf{1.19}$ & $\mathbf{69.7}$ &  $\mathbf{15.4}$ \\
		\hline
		Males & 5.30 &  5.15 & 3.00 &  1.45 & 66.4 &  16.4 \\
		\hline
		Females & 3.51 &  1.06 & 2.55 &  0.36 & 75.5 &  12.2 \\
		\hline
		\hline
		\textbf{Control} & $\mathbf{3.52}$ &  $\mathbf{0.91}$ & $\mathbf{2.76}$ &  $\mathbf{0.44}$ & $\mathbf{80.3}$ &  $\mathbf{9.7}$ \\
		\hline
		Males & 3.77 &  1.05 & 2.84 &  0.51 & 77.5 &  10.3 \\
		\hline
		Females & 3.10 &  0.36 & 2.62 &  0.22 & 85.0 &  6.5 \\
		\hline
	\end{tabular}
	}
	\caption{distribution of reading task durations and phonation times. Reading task duration is measured as the time difference between the beginning of the first word of the phrase and the ending of the last word. Phonation time is measured as the portion of the reading task during which the reader is voicing. Phonation time is given both in absolute units (seconds), and relative to the duration of the reading task.}
	\label{tab:Times}
\end{table}

\section{Methods}
\label{sec:Methods}
\subsection{Voice Analysis}
\label{subsec:VoiceAnalysis}
The recorded voice signals were band-pass filtered to discard all spectral energy outside the microphone bandwidth (30 Hz to 18000 Hz). This filtering was performed using the discrete Fourier transform (DFT) with previous zero padding \cite[][chap.7]{MaIn11}. The fundamental frequency contour of each filtered signal was estimated using the YIN algorithm \cite{ChKa02}, using an integration window length of 16.7 ms. The YIN algorithm is based on the computation of the autocorrelation function. It provided a sequence of fundamental frequency estimates $f_\mathrm{o}\left[n\right]$  at a sampling rate equal to the sampling rate of the signal divided by 32 (approx. 1378 Hz). A null value was assigned to the fundamental frequency for unvoiced samples, which corresponded to silent pauses. Afterwards, all estimates were manually revised and corrected by visually comparing them to the first harmonic of the spectrogram.

The range of the fundamental frequency values corresponding to each signal was described by the following parameters, extracted from the $f_\mathrm{o}\left[n\right]$ sequence:
\begin{itemize}
	\item Mean ($\mu_{f_\mathrm{o}}$):
	\begin{equation}
	\mu_{f_\mathrm{o}} = \frac{1}{N} \sum_n f_\mathrm{o}\left[n\right]
	\end{equation}
	where $N$ is the number of non-null samples of $f_\mathrm{o}\left[n\right]$:
	\begin{equation}
	N = \sum_n v\left[n\right]; \ \ v\left[n\right] = \left\lbrace \begin{array}{ll}
	1 & \mathrm{if} \ f_\mathrm{o}\left[n\right]\ne 0 \\
	0 & \mathrm{if} \ f_\mathrm{o}\left[n\right]= 0
	\end{array} \right.
	\end{equation}
	\item Minimum ($f_\mathrm{omin}$):
	\begin{equation}
	f_\mathrm{omin} = \min_{v\left[n\right]=1}\left\lbrace f_\mathrm{o}\left[n\right]\right\rbrace
	\end{equation}
	\item Maximum ($f_\mathrm{omax}$):
	\begin{equation}
	f_\mathrm{omax} = \max_{v\left[n\right]=1}\left\lbrace f_\mathrm{o}\left[n\right]\right\rbrace
	\end{equation}
	\item Standard deviation ($\sigma_{f_\mathrm{o}}$):
	\begin{equation}
	\sigma_{f_\mathrm{o}} = \sqrt{\frac{\sum_n v\left[ n \right] \left( f_\mathrm{o} \left[ n \right] - \mu_{f_\mathrm{o}} \right) ^2}{N}}
	\end{equation}
\end{itemize}

In addition to the previous parameters, all extracted from the time-domain $f_\mathrm{o}\left[n\right]$ sequence, the modulation spectrum of the fundamental frequency contour corresponding to each speaker was estimated from the DFT of its autocorrelation function $\rho_{f_\mathrm{o}}\left[m\right]$. This was defined as:
\begin{equation}
\rho_{f_\mathrm{o}}\left[m\right] = \frac{\sum_n v\left[n\right]v\left[n+m\right]\hat{f}_\mathrm{o}\left[n\right]\hat{f}_\mathrm{o}\left[n+m\right]}{N\sigma_{f_\mathrm{o}}^2}
\label{eq:Autocorrelation}
\end{equation}
where $\hat{f}_\mathrm{o}\left[n\right] = f_\mathrm{o}\left[n\right] - \mu_{f_\mathrm{o}}$. The estimated power spectral density (PSD) for each fundamental frequency contour was:
\begin{equation}
\mathcal{P}_{f_\mathrm{o}}\left(f_k\right) = \sum_{m=-M}^M \rho_{f_\mathrm{o}}\left[m\right]\mathrm{e}^{-j\frac{2\pi k m}{2M+1}}
\end{equation}
where $f_k = \frac{k}{2M+1}\cdot f_\mathrm{s}$, with $-M \le k \le M$; $f_\mathrm{s}$ is the sampling frequency of the fundamental frequency contour ($f_\mathrm{s} \approx 1378 \ \mathrm{Hz}$); and $M$ is the maximum value of the time lag $m$ in the estimated autocorrelation function. The value of $M$ was set to $M=2f_\mathrm{s}$, so the PSD was estimated with a resolution of approximately 0.25 Hz. The average PSD within a frequency interval $\left(f_\mathrm{a}, f_\mathrm{b}\right]$, referred to as $\overline{\mathcal{P}}_{f_\mathrm{o}}\left(f_\mathrm{a}, f_\mathrm{b}\right)$, was estimated as the average value of $\left| \mathcal{P}_{f_\mathrm{o}}\left(f_k\right)\right|$ for $f_\mathrm{a} < f_k \le f_\mathrm{b}$. From the average PSD estimated for different bands, the following parameters were calculated:
\begin{itemize}
	\item The low-frequency energy ratio ($LFER$):
	\begin{equation}
		LFER = \frac{\overline{\mathcal{P}}_{f_\mathrm{o}}\left(0, 6\right) \cdot \left(6-0\right)}{\overline{\mathcal{P}}_{f_\mathrm{o}}\left(0, 20\right) \cdot \left(20-0\right)}
		\label{eq:LFER}
	\end{equation}
	\item The mid-frequency energy ratio ($MFER$):
	\begin{equation}
	MFER = \frac{\overline{\mathcal{P}}_{f_\mathrm{o}}\left(6, 12\right) \cdot \left(12-6\right)}{\overline{\mathcal{P}}_{f_\mathrm{o}}\left(0, 20\right) \cdot \left(20-0\right)}
	\label{eq:MFER}
	\end{equation}
	\item The high-frequency energy ratio ($HFER$):
	\begin{equation}
	HFER = \frac{\overline{\mathcal{P}}_{f_\mathrm{o}}\left(12, 20\right) \cdot \left(20-12\right)}{\overline{\mathcal{P}}_{f_\mathrm{o}}\left(0, 20\right) \cdot \left(20-0\right)}
	\label{eq:HFER}
	\end{equation}
\end{itemize}

\subsection{Statistical Methods}
\subsubsection{Correlation}
The correlation between the afore-defined parameters describing the fundamental frequency contours, and the PD stage of speakers, as assessed by the H\&Y scale, has been evaluated  using the Spearman coefficient $\rho_\mathrm{s}$ \cite{Spea04} with correction for tied observations \citep[][p. 429]{GiCh03}. This non-parametric measure of correlation was preferred to the Pearson coefficient due to the limited number of samples. Correlations between the same parameters and the age of the speakers have also been evaluated for comparison purposes. The $p$-value indicating the probability of two independent random samples having a Spearman correlation coefficient $\rho_\mathrm{s}$ higher than a given value has been estimated by assuming a standardised normal distribution of the following statistic \citep[][p. 428]{GiCh03}:
\begin{equation}
Z = \rho_\mathrm{s} \sqrt{\mathcal{N}-1}
\end{equation}
where $\mathcal{N}$ is the number of voice samples ($\mathcal{N}=62$ in this case).

\subsubsection{Regression}
Additionally, the capability of intonation descriptors to model progression of the PD has been assessed by linear regression. Specifically, some of the parameters defined before have been used as independent variables to build linear regression models for H\&Y labels. In order to make the distributions of such parameters closer to Gaussian, their natural logarithms were calculated as a preprocessing step before regression. For instance, the regression model built with $LFER$ and $MFER$ would be:
\begin{equation}
\hat{HY} = r_0 + r_1 \cdot \log LFER + r_2 \cdot \log MFER
\end{equation}
where $\hat{HY}$ is the H\&Y label estimated by the model, and $r_0$, $r_1$ and $r_2$ are the regression coefficients. The regression coefficients were estimated using the normal equations \cite[][chap.10]{Renc02} for mean square error minimisation. The quality of each linear model was evaluated by means of the coefficient of determination $R^2$, calculated as:
\begin{equation}
R^2 = 1- \frac{\mathrm{Var}\left[\hat{HY} - HY \right]}{\mathrm{Var}\left[HY \right]}
\end{equation}
where $\mathrm{Var}\left[\cdot\right]$ means variance, and $HY$ are the actual H\&Y labels.

\subsubsection{Test of means}
The performance of the regression model was evaluated by means of the coefficient of determination $R^2$ mentioned before, and also analysing the distribution of $\hat{HY}$ values assigned to the speakers in each of the groups defined by their actual H\&Y labels, i.e. five groups defined by labels 0 (healthy speakers), 1, 2, 3 and 4. The hypotheses of the $\hat{HY}$ distributions of two of such groups having different means was evaluated using the non-parametric Wilcoxon test \cite[][chap. 2]{Higg04}.

\section{Results}
\subsection{Correlation analysis}
Table \ref{tab:Correlations} shows the values of the Spearman coefficients $\rho_\mathrm{s}$ obtained when evaluating correlations between all the parameters defined in section \ref{subsec:VoiceAnalysis} and the H\&Y label assigned to each speaker, and also between the same parameters and the age of the speakers. These results indicate that there is a significant positive relation of the mean fundamental frequency of the male speakers with their age, which corresponds to equally significant increases of the minimum and maximum $f_\mathrm{o}$ with age. Thus, an overall positive shift in the fundamental frequency range as a function of age is detected for the recorded male population. However, in what refers to PD progression correlations are less relevant. In the case of males, only a significant increase of $f_\mathrm{omin}$ with H\&Y label is detected. This corresponds to an almost significant increase of $f_\mathrm{omin}$ with H\&Y label for females too, which results in a significant correlation for the overall population. For men, the significant effect of PD on the increase of $f_\mathrm{omin}$ is associated to an almost significant increase in the mean fundamental frequency $\mu_{f_\mathrm{o}}$. For women, the almost significant increase in $f_\mathrm{omin}$ is coincident with an almost significant reduction in $f_\mathrm{omax}$, which results in a significant reduction of the pitch range, as measured by $\sigma_{f_\mathrm{o}}$.

\begin{table}
	\centering
	{\scriptsize \renewcommand{\arraystretch}{1.5}
		\begin{tabular}{|@{\hskip2pt}r@{\hskip2pt}||@{\hskip2pt}c@{\hskip2pt}|@{\hskip2pt}c@{\hskip2pt}|@{\hskip2pt}c@{\hskip2pt}||@{\hskip2pt}c@{\hskip2pt}|@{\hskip2pt}c@{\hskip2pt}|@{\hskip2pt}c@{\hskip2pt}|}
			\hline
			&  \textbf{Age} & \textbf{Age} &  \textbf{Age} &  \textbf{H\&Y} &  \textbf{H\&Y} &  \textbf{H\&Y}\\
			& \textbf{(men)} & \textbf{(women)} & \textbf{(all)} & \textbf{(men)} & \textbf{(women)} & \textbf{(all)}\\
			\hline
			\hline
			$\mu_{f_\mathrm{o}}$ & \textbf{0.62} ($< 0.01$) & 0.00 ($0.49$) & 0.31 ($0.01$) & 0.33 ($0.02$) & 0.10 ($0.31$) & 0.20 ($0.05$) \\
			\hline
			$f_\mathrm{omin}$ & \textbf{0.44} ($< 0.01$) & -0.15 ($0.25$) & 0.21 ($0.05$) & \textbf{0.46} ($<0.01$) & 0.47 ($0.01$) & \textbf{0.40} ($<0.01$) \\
			\hline
			$f_\mathrm{omax}$ & \textbf{0.53} ($< 0.01$) & -0.22 ($0.18$) & 0.21 ($0.05$) & 0.17 ($0.13$) & -0.43 ($0.02$) & -0.04 ($0.38$) \\
			\hline
			\hline
			$\sigma_{f_\mathrm{o}}$ & 0.33 ($0.02$) & -0.10 ($0.32$) & 0.12 ($0.19$) & -0.24 ($0.08$) & \textbf{-0.64} ($<0.01$) & \textbf{-0.31} ($<0.01$) \\
			\hline
			$\frac{\sigma_{f_\mathrm{o}}}{\mu_{f_\mathrm{o}}}$ & -0.07 ($0.35$) & -0.15 ($0.23$) & -0.10 ($0.23$) & \textbf{-0.52} ($<0.01$) & \textbf{-0.76} ($<0.01$) & \textbf{-0.58} ($<0.01$) \\
			\hline
			\hline
			$LFER$ & -0.06 ($0.35$) & -0.03 ($0.47$) & -0.02 ($0.45$) & \textbf{-0.42} ($<0.01$) & \textbf{-0.50} ($<0.01$) & \textbf{-0.43} ($<0.01$) \\
			\hline
			$MFER$ & 0.03 ($0.43$) & -0.03 ($0.47$) & -0.01 ($0.48$) & \textbf{0.39} ($<0.01$) & 0.46 ($0.01$) & \textbf{0.42} ($<0.01$) \\
			\hline
			$HFER$ & 0.23 ($0.08$) & 0.06 ($0.40$) & 0.13 ($0.16$) & \textbf{0.50} ($<0.01$) & 0.12 ($0.31$) & \textbf{0.34} ($<0.01$) \\
			\hline
		\end{tabular}
	}
	\caption{Spearman correlation coefficients between the signal parameters defined in section \ref{subsec:VoiceAnalysis}, speaker age and assigned H\&Y labels. Correlations have been evaluated for men and women separately, and also jointly. Estimated $p$-values are indicated between brackets. Significant correlations at the 99\% significance level ($p<0.01$) are highlighted in bold.}
	\label{tab:Correlations}
\end{table}

Regarding the fundamental frequency range, two observations can be made about the results in Table \ref{tab:Correlations}. The first one is that significant correlations between $\sigma_{f_\mathrm{o}}$ and PD progression have been found, while this is not the case for age. Therefore, $\sigma_{f_\mathrm{o}}$ is a more specific marker of PD than $f_\mathrm{omin}$. The second observation is that the magnitude and significance of correlations between fundamental frequency range and H\&Y labels can be increased by making the measurement of range relative to the mean fundamental frequency, that is, measuring the range as $\frac{\sigma_{f_\mathrm{o}}}{\mu_{f_\mathrm{o}}}$. Significant correlations between this parameter and H\&Y labels have been found for both men and women, and for the overall population. For these three groups, the absolute value of the correlation coefficients are above 0.5.

The normalised standard deviation of $f_\mathrm{o}$, $\frac{\sigma_{f_\mathrm{o}}}{\mu_{f_\mathrm{o}}}$, provides a description of the range of variation of $f_\mathrm{o}$, that is, the amplitude of the fundamental frequency contour. Yet, it does not contain information on the rate of variation of the fundamental frequency. i.e. its modulation rate or modulation frequency. This can be obtained from the Fourier transform of the normalised autocorrelation function (\ref{eq:Autocorrelation}), which is an estimate of the PSD of $f_\mathrm{o}\left[n\right]$, as explained before. In order to analyse the PSD, three modulation frequency intervals were defined as explained next. 

The sentence read by the speakers, mentioned in section \ref{sec:Materials}, consists of 14 syllables. Reading it took 2.35 seconds to the fastest reader; therefore, modulation frequencies above $\frac{14}{2.35}\approx 5.96 \ \mathrm{Hz}$ approximately correspond to $f_\mathrm{o}$ modulations at intra-syllabic level for all the speakers. Correspondingly, $f_\mathrm{o}$ modulations at supra-syllabic level are expected to happen at rates below 5.96 Hz for the faster speakers, and at even lower rates for the rest. In addition, it is known that frequency modulations at rates above approximately 12 Hz cannot be tracked as $f_\mathrm{o}$ variations by the ear; instead they are perceived as roughness \cite{Dani08}. Last, some of the most usual auditory models include filters performing time integration with integration times in the range of tens of milliseconds \cite{PaHo96}. This corresponds to bandwidths in the range of a few tens of hertzs, beyond which there is little sense in speaking about modulations. Considering these references, three modulation frequency bands have been defined:
\begin{itemize}
	\item A low-frequency band corresponding to modulations at supra-syllabic level: 0 to 6 Hz.
	\item A mid-frequency band corresponding to modulations at intra-syllabic level: 6 to 12 Hz.
	\item A high-frequency band corresponding to modulations perceived as rough\-ness: 12 to 20 Hz.
\end{itemize}
These three bands were considered for calculating the $LFER$, $MFER$ and $HFER$ parameters defined in equations (\ref{eq:LFER}) to (\ref{eq:HFER}). The results of evaluating the correlations between these three parameters and both age and H\&Y label are presented in the three bottom rows of Table \ref{tab:Correlations}. No significant correlations have been detected with age. By contrast, the $LFER$ parameter is significantly reduced for PD patients, as indicated by the moderate and significant negative correlations between $LFER$ and H\&Y label. This reduction of modulation energy at low frequencies corresponds to increases at both mid frequencies and high frequencies. Such fact indicates that the fundamental frequency changes faster for speakers suffering from PD at more advanced stages. For the male population, the increase in modulation energies is more relevant for high frequencies, i.e. voices tend to be rougher; for females, such increase is more relevant for mid frequencies, which means more variations of fundamental frequency at intra-syllabic level.

\subsection{Predictive analysis} 

\begin{figure}
	\centering
	\includegraphics[width = 13.5cm]{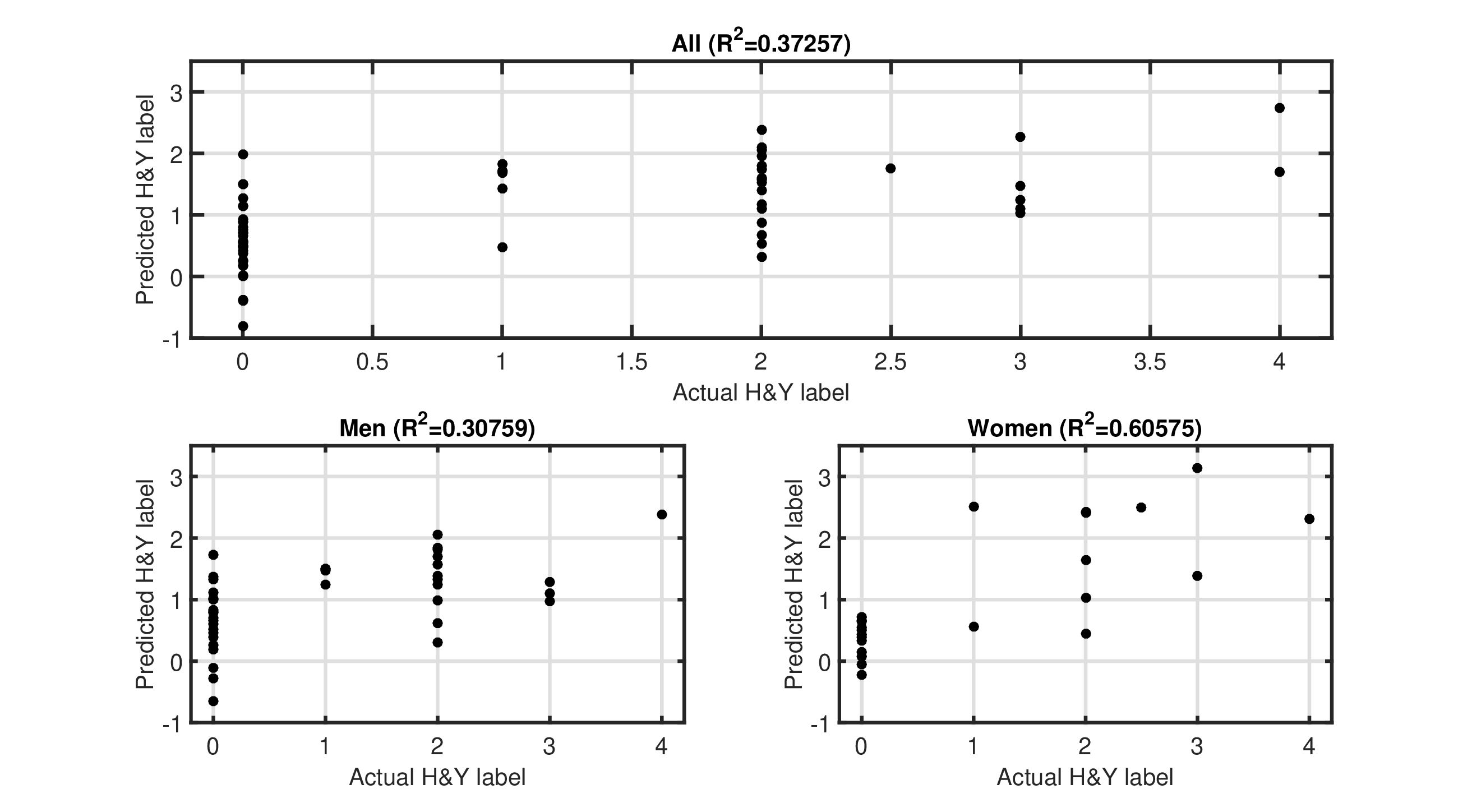}
	\caption{Scatter plots showing the relation between actual H\&Y labels and those by the regression model for the overall population being studied (upper plot), and differentiated for men (lower left plot) and women (lower right). The variables included in the regression model were $LFER$, $MFER$, and $\frac{\sigma_{f_\mathrm{o}}}{\mu_{f_\mathrm{o}}}$.}
	\label{fig:Regression}
\end{figure}

The three parameters in Table \ref{tab:Correlations} yielding the highest correlation coefficients with PD progression were selected for building a regression model for H\&Y labels: $LFER$, $MFER$ and $\frac{\sigma_{f_\mathrm{o}}}{\mu_{f_\mathrm{o}}}$. $HFER$ was discarded for carrying redundant information once $LFER$ and $MFER$ were considered. Indeed, all these three parameters add up to one, since each of them represents the fraction of modulation energy in a specific sub-band of the 0-20 Hz range, and they cover this frequency range fully and without overlap. $\sigma_{f_\mathrm{o}}$  was not selected since it measures the same intonation feature as $\frac{\sigma_{f_\mathrm{o}}}{\mu_{f_\mathrm{o}}}$, i.e. the fundamental frequency range. Last, $f_\mathrm{omin}$ was not considered because for the population being analysed it was not a specific indicator of PD, given its relevant correlation with age. The modelling results are depicted in Figure \ref{fig:Regression}. The value of the corresponding determination coefficient was $R^2 \approx 0.373$ when all the speakers were modelled as a single population. A relevant difference was found in the quality of the modelling results when the modelling was applied only to male ($R^2 \approx 0.308$) or to female speakers ($R^2 \approx 0.606$).

The graph in Figure \ref{fig:Regression} qualitatively shows that while there is a general direct relation between the actual H\&Y labels associated to the speakers (0 for the control individuals, and the label assigned by the neurologist for the PD patients) and the labels assigned by the regression model, such relation is mainly due to the difference between the control speakers and the rest. In other words, the relation would be less evident if the control speakers were left out of the analysis. The relevance of this observation can be confirmed by evaluating differences in the mean value of the modelling results corresponding to the five speaker groups: control (H\&Y label equal to 0), and PD patients with assigned labels 1, 2, 3, and 4. The patient with label 2.5 was left out of this analysis for being the only one in its group. Table \ref{tab:Wilcoxon} shows the $p$-values evaluating the differences among these means according to the Wilcoxon test. These confirm that the only group significantly differentiated from the rest is the control group. As regards statistical significance, the same results were found for male and female speakers when modelled separately.

\begin{table}
\centering
{\footnotesize
\begin{tabular}{|r|cccc|}
\hline
\textbf{H\&Y label} & 1 & 2 & 3 & 4\\
\hline
0 & \textbf{$<$0.01} & \textbf{$<$0.01} & \textbf{$<$0.01} & \textbf{$<$0.01}\\
1 & & 0.46 & 0.27 & 0.21\\
2 & & & 0.42 & 0.10\\
3 & & & & 0.10\\
\hline
\end{tabular}
}
\caption{$p$-values for the Wilcoxon test evaluating differences on the distributions of H\&Y labels assigned by the regression model to the different speaker groups. Control speakers correspond to group 0.}
	\label{tab:Wilcoxon}
\end{table}

\begin{figure}
	\centering
	\includegraphics[width = 13cm]{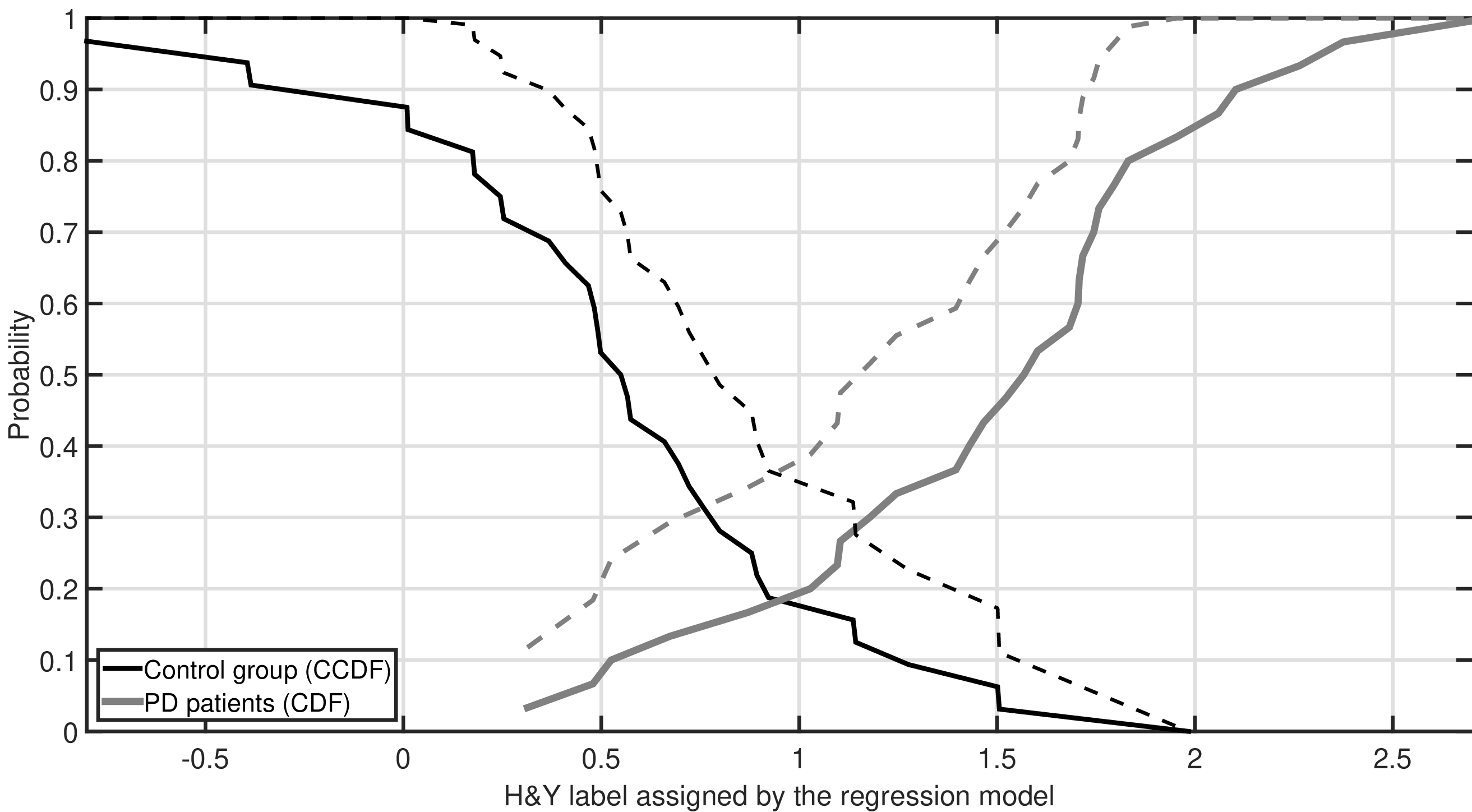}
	\caption{Experimental CDF of the H\&Y labels produced by the regression model for PD patients (grey); and complementary CDF of labels produced by the same model for the control group (black). The dashed lines indicate the 99\% confidence intervals, calculated as proposed in \cite[][p. 15]{Higg04}.}
	\label{fig:ModelCDF}
\end{figure}

As for the question of the feasibility of using these three intonation descriptors (i.e. $LFER$, $MFER$, and $\frac{\sigma_{f_\mathrm{o}}}{\mu_{f_\mathrm{o}}}$) as cues for the early diagnosis of PD, the regression model described before may be potentially used for discriminating between PD patients and healthy speakers. This could be achieved by setting a threshold in the model output $\hat{HY}$. A first assessment of the performance of such a system can be obtained by analysing the distributions of $\hat{HY}$ for both PD patients and healthy controls. The plot in Figure \ref{fig:ModelCDF} shows the cumulative distribution function (CDF) of the model outputs $\hat{HY}$ for PD patients, and the complementary CDF (CCDF) for healthy individuals. The model was the same one that produced the results in Figure \ref{fig:Regression}. The crossing point between the CDF corresponding to PD patients and the CCDF corresponding to control speakers indicates an estimated value for the equal error rate ($EER$) of PD detection based on this model of 18\%. If the 99\% confidence intervals for the sample CDFs are considered, the $EER$ estimate is increased up to 36\%.

\begin{figure}[t]
	\centering
	\includegraphics[width = 11cm]{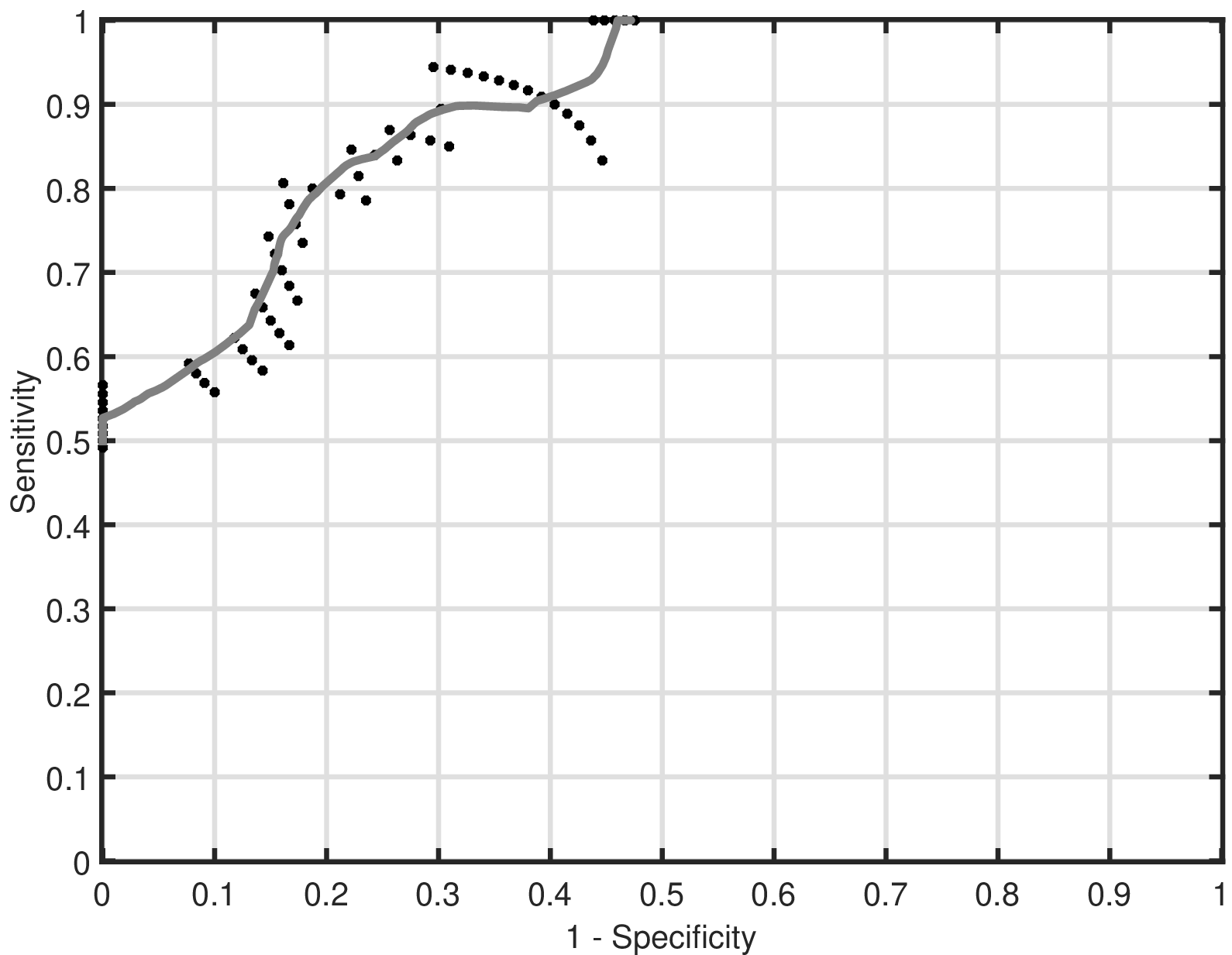}
	\caption{ROC curve corresponding to the detection of PD using the regression model mentioned in the text. The points correspond to the actual results obtained with the dataset. The grey line corresponds to a local averaging of the points. The area under the line equals 0.634, which corresponds to an estimation of the $AUC$.}
	\label{fig:ROC}
\end{figure}

The relation between specificity and sensitivity of this detection system is represented in the receiver operating characteristic (ROC) plot in Figure \ref{fig:ROC}. The points in this graph correspond to the (1-\emph{specificity}, \emph{sensitivity}) pairs associated to the same threshold values for the detection system as the H\&Y labels in the horizontal axis of Figure \ref{fig:ModelCDF}. The continuous line results from smoothing the point data by local averaging, and it provides an estimate of the ROC curve of the system. The area under this curve ($AUC$) provides a standard measure of system performance \cite{Brad97}. Its value for this system is 0.634.

\section{Discussion}

The fact that PD affects pitch range in intonation has been pointed out by several researchers \cite{RCRR11,SGMS13,INSH15}. Specifically, a reduced pitch range has been measured in parkinsonian voices for the speakers performing diverse tasks: text reading \cite{RCRR11,SGMS13}, monologue \cite{RCRR11}, and conversation \cite{INSH15}. However, Bandini \emph{et al} did not find relevant differences in pitch range between PD patients and healthy speakers in repeated sentences \cite{BGOB15}. In this paper, we have presented a transversal study of intonation in parkinsonian voices performing a reading task. Accordingly, the results should be comparable to those obtained from reading tasks in \cite{RCRR11,SGMS13}.

The first step of our analysis has been to obtain the fundamental frequency contours corresponding to every speaker. These contours have been described by the mean $\left(\mu_{f_\mathrm{o}}\right)$, maximum $\left(f_\mathrm{omin}\right)$ and minimum values $\left(f_\mathrm{omax}\right)$ of the fundamental frequency contour $\left(f_\mathrm{o}\left[n\right]\right)$. The results in Table \ref{tab:Correlations} show that these three parameters are positively and significantly correlated with age for male speakers, and that only $f_\mathrm{omin}$ has a relevant correlation with PD stage, significant for males and almost significant for females. The values of the correlation coefficients between $f_\mathrm{omin}$  and age, and between $f_\mathrm{omin}$ and H\&Y label are very similar. The fact that the fundamental frequency tends to increase with age for men over 60 years old has been widely reported in the scientific literature \cite{ScMu07,KMGS17}. For women, the trend is the opposite (the fundamental frequency tends to decrease) and less pronounced \cite{KMGS17}. The results in Table \ref{tab:Correlations} are consistent with this previous knowledge: moderate and significant positive correlations between fundamental frequency and age for males, and low negative correlations for women not reaching statistical significance for this population. As for correlation with PD progression, only $f_\mathrm{omin}$ exhibits a significant correlation with H\&Y labels in the case of men. Since the value of this correlation is similar to the correlation with age, it can be concluded that the relation between $f_\mathrm{omin}$ increase and PD is not more relevant than its relation with age. By contrast, for women PD has an effect on $f_\mathrm{omin}$ that is opposite to that of age, with the correlation between $f_\mathrm{omin}$ and H\&Y labels almost reaching statistical significance. This increase of $f_\mathrm{omin}$ as PD progresses is coincident with a similar trend of reduction in $f_\mathrm{omax}$. These sex-related differences in the prosodic disorders experienced by parkinsonian speakers are in agreement with previous findings reported in the scientific literature \cite{MaHS11,SkVS11}.

The simultaneous positive correlation between  $f_\mathrm{omin}$ and H\&Y labels, and negative correlation between $f_\mathrm{omax}$ and H\&Y labels for female speakers indicates a relation between PD and a reduction of pitch range. This is concordant with the usual identification of monotony as a feature of parkinsonian voices \cite{Jank08}. Skodda \emph{et al} also found that the perceived monotony tended to increase with time for PD patients \cite{SGMS13}. Pitch range has been evaluated in the literature both as the difference between $f_\mathrm{omax}$ and $f_\mathrm{omin}$ (e.g. \cite{BGOB15,INSH15}) and the standard deviation of $f_\mathrm{o}\left[n\right]$, $\sigma_{f_\mathrm{o}}$ (e.g. \cite{RCRR11}). We have preferred this second option. The results in Table \ref{tab:Correlations} indicate that the relation between $\sigma_{f_\mathrm{o}}$ and PD progression is notably different to the relation between $\sigma_{f_\mathrm{o}}$ and age. Thus, the analysis of $\sigma_{f_\mathrm{o}}$ is expected to provide information more specific to PD than $\mu_{f_\mathrm{o}}$, $f_\mathrm{omin}$, or $f_\mathrm{omax}$. As for the specific values of the calculated correlation coefficients, $\sigma_{f_\mathrm{o}}$ only has a significant negative correlation with H\&Y labels for women. For the overall population, the correlation reaches statistical significance but its absolute value is low (0.31). This result is congruent with the lack of relevance of the differences between PD patients and healthy controls in the absolute pitch range informed by Bandini \emph{et al} \cite{BGOB15}. On the opposite, the significant results reported by Rusz \emph{et al} \cite{RCRR11} and by Ikui \emph{et al} \cite{INSH15} were obtained with relative measurements of pitch range. For our measurements, dividing $\sigma_{f_\mathrm{o}}$ by $\mu_{f_\mathrm{o}}$ provides a relative evaluation of pitch range more correlated to PD progression. In fact, significant correlations are obtained for both male and female speakers, and the values of the correlation coefficients are all over 0.5. 

\begin{figure}[t]
	\centering
	\includegraphics[width = 13cm]{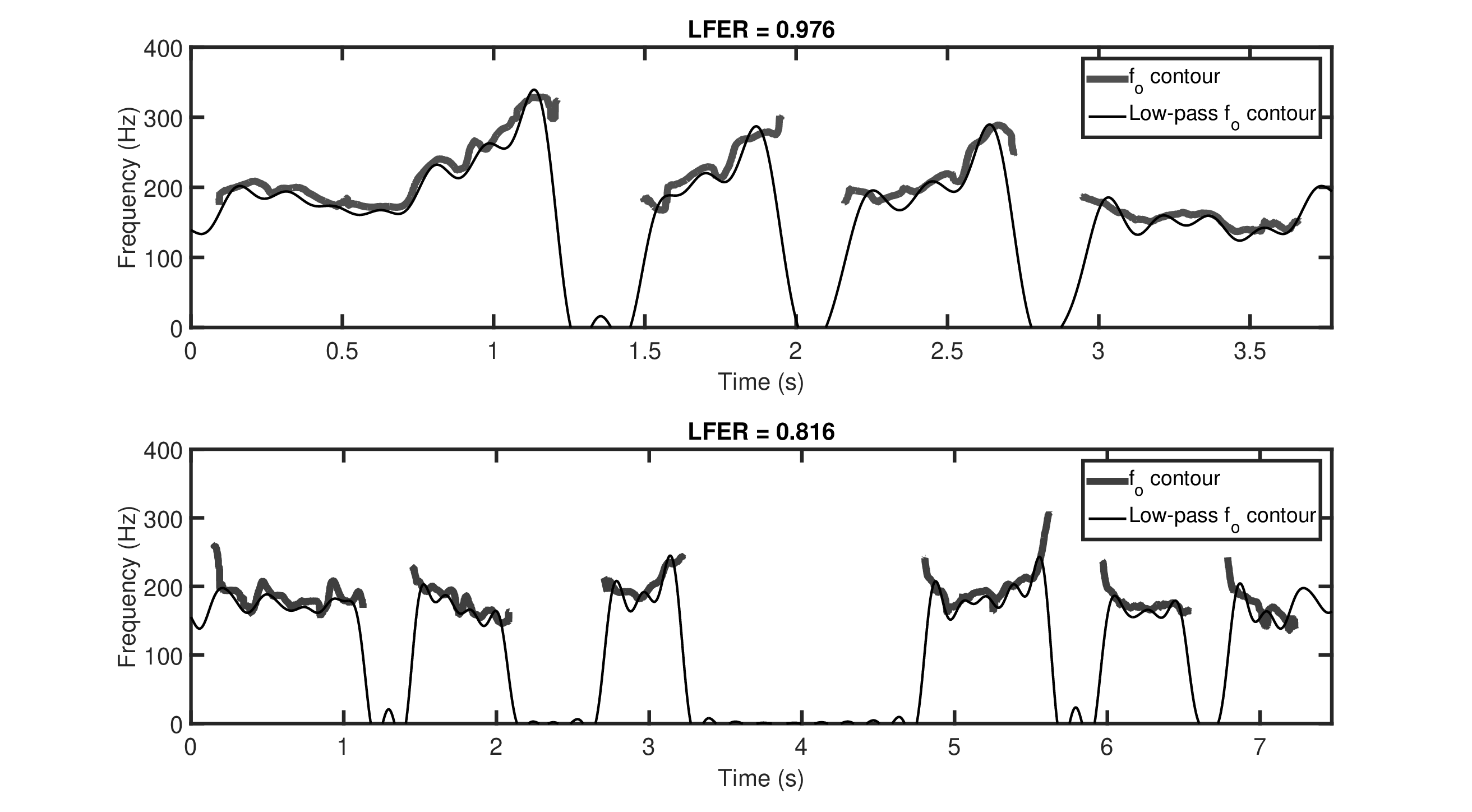}
	\caption{Fundamental frequency contours (thick lines) and their components with modulation frequencies up to 6 Hz (thin lines) for two voices with different values of the $LFER$. The low-frequency components were estimated by calculating the DFT of the fundamental frequency contour, zeroing all its components corresponding to frequencies above 6 Hz, and going back to temporal domain by computing the inverse DFT. The non-voiced intervals were managed by using the DFT for non-uniformly spaced samples instead of the standard DFT, as in \cite{FSOG15}.}
	\label{fig:PitchContours}
\end{figure}

The relative pitch range $\frac{\sigma_{f_\mathrm{o}}}{\mu_{f_\mathrm{o}}}$ provides information about the magnitude of pitch variations, but not about the speed, or rate, at which such variations happen. In order to quantify changes on that pitch modulation rate, we defined the parameters $LFER$, $MFER$, and $HFER$ (equations (\ref{eq:LFER}) to (\ref{eq:HFER})). The values of the correlation coefficients in Table \ref{tab:Correlations} show that the reduction of pitch range associated to PD is accompanied by a significant reduction in $LFER$. This implies that the modulation rate of fundamental frequency, or pitch, increases as PD progresses. Thus, for the studied population pitch changes less and more rapidly as the speakers suffer from PD at more advanced stages of the illness. This observation is consistent with the findings of Cnockaert \emph{et al} \cite{CSAO08} relative to the higher modulation frequencies observed in PD patients with respect to healthy speakers.  Figure \ref{fig:PitchContours} illustrates this effect. It shows the fundamental frequency contours corresponding to two speakers with $LFER$ values equal to 0.976 and 0.816, respectively. Four voiced segments can be clearly identified in the upper contour, while the lower one is formed by six voiced segments. For the upper plot (higher $LFER$) the low-frequency components of the fundamental frequency contour provide a good description of the overall evolution of $f_\mathrm{o}$. The intonation of the three first voiced segments has a repetitive pattern that contributes to increasing the low-frequency component of $f_\mathrm{o}$ modulation. On the other hand, the voiced segments in the lower plot (lower $LFER$) have intonation patterns more independent from one another. This, together with the presence of some rapid variations in $f_\mathrm{o}$, especially in the first two voiced segments, contributes to lessen the value of $LFER$. One additional trait of the intonation contour in the lower plot of Figure \ref{fig:PitchContours} is the presence of some rapid $f_\mathrm{o}$ decays at the beginning of several voiced segments, and one rapid increase at the end of the fourth voiced segment. These fast variations of $f_\mathrm{o}$ at the beginnings and ends of voiced segments contribute to the decrease of the $LFER$, and they are common in PD patients \cite{MaHS11}.

The results depicted in Figure \ref{fig:Regression} suggest that estimating the PD stage from only the relative fundamental frequency range, $LFER$ and $MFER$ is far from being feasible. The combined used of the logarithms of these three parameters in a linear regression model is able to explain only 37\% of the variance of H\&Y labels ($R^2 \approx 0.373$). This result is in agreement with the limited value of the correlation coefficients reported in Table \ref{tab:Correlations}, with all of them having absolute values below 0.8, and those referring to the overall population having absolute values below 0.6. The stratified analysis of the modelling results considering the actual H\&Y labels assigned by the neurologist leads to the conclusion that the linear regression model only provides significant differences between the healthy speakers and the rest, but it is not able to significantly discriminate the H\&Y label assigned to each patient by the neurologist (Table \ref{tab:Wilcoxon}). Although the regression results are better for women when different regression models are produced for male and female speakers (Figure \ref{fig:Regression}), the limited number of women in the studied population does not allow to get significant discrimination among PD stages either. Rektorova \emph{et al} \cite{RMJK16} found pitch range to be a good predictor of the cognitive status of PD patients at a given time, but not of the cognitive decline experienced after a certain period. 

One last question regarding the usefulness of this intonation analysis is whether it may be used as a tool for the early diagnosis of PD. Figure \ref{fig:ModelCDF} shows the distribution of H\&Y labels provided by the same regression model mentioned before. Due to the limited number of samples in the experiment, the 99\% confidence interval for the distribution has also been plotted. The crossing point between the distribution of labels for the control group and the complementary distribution of labels for patients provides an estimate of the potential performance of the detection system when it is calibrated so that the false positive rate equals the false negative rate. This crossing point for 99\% confidence intervals for the distributions is at $EER=36.2\%$. The potential performance of such a system can also be evaluated by plotting the ROC curve, which represents the relation between specificity and sensitivity (Figure \ref{fig:ROC}). Again, due to the limited number of samples, merely linking the points in the plot would produce a zigzagging curve. For this reason, we have opted for performing a local averaging of points before generating the ROC curve. Its corresponding $AUC$ results in a value similar to $1-EER$, $AUC=0.634$.

The performance indicators obtained for the detection system mentioned above may be evaluated from two different points of view. On the one hand, the values for $1-EER$ and $AUC$ are well over 0.5, which implies that this simple model including only three parameters contains significant information for the detection of PD. The fact that PD has been reported to affect cognition previously to motion \cite{Adle11} and that cognition affects intonation planning \cite{MaHS11} contributes to justify why our regression model partly explains the differences between healthy and parkinsonian speakers, although it is not able to further discriminate the stage of the disease. Perceived monotony has been reported to be correlated with the cognitive decline experienced by PD patients in a longitudinal study \cite{SGMS13}. However, the intrinsic characteristics of a longitudinal study imply that the same population is analysed at two different times, while in transversal studies the patient populations suffering from PD at different stages are also different, so there is an inter-speaker variability among PD stages that does not happen in longitudinal studies. In this study, this inter-speaker variability may have also been affected by the diversity in the state of health of the participants, which was only controlled regarding neurological condition. Specifically, the presence of mild laryngeal disorders was not evaluated, it was only checked that the participants could read intelligibly without excessive effort and without feeling pain. The inter-speaker diversity is a likely cause for this study not resulting in the same significant correlations as reported in \cite{SGMS13}.

On the other hand, while $1-EER$ and $AUC$ have values well over 0.5, these values are also far from the optimum, which is 1. It should be further considered that these performance estimates are optimistic, since they were obtaining using the resubstitution method for evaluation \cite{ThKo03}, although the high ratio between voice samples (62) and number of parameters (3) suggests that this optimistic bias is not too high. Consequently, the value of these parameters for the detection of PD is limited, since their foreseeable performance in the detection of PD is not high, a potential explanation being that inter-speaker variability makes it difficult to set thresholds for the objective measurements of voice monotony that are universally valid for evaluating PD evolution. However, the reported results have also shown that some simple descriptors of the variability of the fundamental frequency are significantly correlated with the H\&Y labels assigned to the patients by the neurologist.

\section{Conclusions}
The analysis of the fundamental frequency contours produced by 30 PD patients plus 32 healthy individuals reading a text with only voiced phonemes in Spanish indicates that the pitch range, when measured relative to the mean fundamental frequency, provides a cue for evaluating PD stage that is fairly independent from age. However, the correlation between pitch range and PD stage only reaches moderate-to-high values in the case of women. The dynamics of the fundamental frequency has been studied not only in terms of its range, but also in
terms of its modulation frequencies. According to the results shown in this paper, this analysis is able to provide significant information for studying the effect of PD on prosody,
since it allows detecting the relevance of the slow  (at supra-syllabic level) components of pitch evolution, which are related to the ability to plan intonation in the long term.
Yet, the quantitative assessment of the performance of the regression models proposed for predicting H\&Y labels as a function of both pitch range and pitch modulation rates suggest that these measures, though informative, they are likely to be of limited value in the early diagnosis of PD, unless they can be combined with other diagnosis cues. Regarding inter-speaker variability, our results are consistent with previous findings pointing out the differences between male and female speakers in the evolution of their intonation patterns as a function of both age and PD disease. This indicates the convenience of developing sex-dependent models for the assessment of intonation in this context.





\section*{Acknowledgements}
This work has been funded by the Spanish government through project grants TEC2012-38630-C04-01 and MAT2015-64139-C4-3-R.

\bibliographystyle{elsarticle-num}
\bibliography{ParkinsonVoice}







\end{document}